\documentstyle[psfig]{mn0}

\begin{document}

\def\simg{\mathrel{%
      \rlap{\raise 0.511ex \hbox{$>$}}{\lower 0.511ex \hbox{$\sim$}}}}
\def\siml{\mathrel{%
      \rlap{\raise 0.511ex \hbox{$<$}}{\lower 0.511ex \hbox{$\sim$}}}}

\parskip 3pt
\onecolumn

\title[Afterglow 090510]{GRB 090510: a short burst from a massive star ?}

%\author[Sue de Nimes]{Ano Neamus \\
\author[A. Panaitescu]{ A. Panaitescu \\
       Space Science and Applications, MS D466, Los Alamos National Laboratory,
       Los Alamos, NM 87545, USA}

\maketitle

\begin{abstract}
\small{\\
 GRB afterglow 090510 is (so far) the best-monitored afterglow in the optical, X-ray, and above 
 100 MeV, measurements covering 2--3 decades in time at each frequency. 
 Owing to its power-law temporal decay and power-law spectrum, it seems very likely that the 
 highest energy emission is from the forward-shock energizing the ambient medium (the standard 
 blast-wave model for GRB afterglows), the GeV flux and its decay rate being consistent with that 
 model's expectations. However, the synchrotron emission from a
 collimated outflow (the standard jet model) has difficulties in accounting for the lower-energy 
 afterglow emission, where a simultaneous break occurs at 2 ks in the optical and X-ray light-curves, 
 but with the optical flux decay (before and after the break) being much slower than in the X-rays 
 (at same time). The measured X-ray and GeV fluxes are incompatible with the higher-energy afterglow
 emission being from same spectral component as the lower-energy afterglow emission, which suggests
 a synchrotron self-Compton model for this afterglow. Cessation of energy injection in the blast-wave 
 and an ambient medium with a wind-like $n \propto r^{-2}$ density can explain all features of the
 optical and X-ray light-curves of GRB afterglow 090510. Such an ambient medium radial structure 
 is incompatible with this short-GRB originating from the merger of two compact stars.
}
\end{abstract}

\section{Introduction}

 The existence of the afterglow emission following Gamma-Ray Bursts (GRBs) has been predicted 
by Paczy\'nski \& Rhoads (1993) and M\'esz\'aros \& Rees (1997a) and has been been monitored in 
the radio (catalog of Frail et al 2003), optical (catalog of Kann et al 2010, Swift-UVOT 
catalog of Oates et al 2009, Roming et al 2009), and X-ray (catalogs: BeppoSAX -- De Pasquale et al 
2006, Swift -- O'Brien et al 2006, Willingale et al 2007) for hundreds of afterglows and, more recently, 
it has been detected above 100 MeV in a few cases.

 The general afterglow picture is that the relativistic outflow that produced the prompt (burst) 
emission interacts with the ambient medium, driving two shocks: one into the circumburst medium 
(the "forward" shock), leading to a progressive deceleration of the blast-wave) and one into the 
GRB ejecta (the "reverse" shock). The shocks accelerate electrons to relativistic energies (up to
at least 10 GeV, in the comoving frame) through the first-order Fermi process or by the 
plasma two-stream instability (Medvedev \& Loeb 1999). Magnetic dissipation in a Poynting 
outflow has also been proposed as the origin of relativistic electrons in GRB outflows (M\'esz\'aros 
\& Rees 1997b, Lyutikov 2006). Turbulence in the shocked fluid or the mentioned Weibel instability 
are believed to be the origin of the $\sim 1$ Gauss magnetic field required to explain the X-ray 
afterglow emission with synchrotron emission from the forward-shock.

 The radio, optical, and X-ray afterglow emission has been identified most often with synchrotron
emission from the ambient medium energized by the forward-shock (e.g. M\'esz\'aros \& Rees 1997a,
Sari, Piran \& Narayan 1997). The reasons for that is that, for an impulsive GRB ejecta release,
most of the ejecta energy is quickly transferred to the forward-shock, and that the forward-shock
emission flux is expected to decay as a power-law for long times (as observed in GRB afterglows)
without any further assumptions/requirements: just the power-law deceleration of the blast-wave 
and the power-law spectrum of the shock-accelerated electrons suffice to yield a synchrotron flux
with a power-law decay in time. 

 The emission from the GRB ejecta energized by the reverse-shock has also been proposed to be
the afterglow source, more likely at optical frequencies (M\'esz\'aros \& Rees 1997a, Panaitescu 
\& M\'esz\'aros 1998), and was identified as the origin of the bright optical counterpart of
GRB 990123 (Sari \& Piran 1999). Due to the fast cooling of the ejecta electrons, the reverse-shock 
emission may account for the afterglow emission lasting for days only if new electrons are 
accelerated at that shock, i.e. only if there is a continuous influx of ejecta crossing the 
reverse-shock. In this case, the afterglow light-curve also depends on the rate at which the 
new ejecta cross the reverse-shock, a fact used by Uhm \& Beloborodov (2007) and by Genet, Daigne 
\& Mochkovitch (2007) to explain the plateaus displayed by the X-ray afterglow light-curves at 1-10 ks. 
As shown in those articles, the reverse-shock emission may also account for the various degree of 
coupling between the optical and X-ray afterglow light-curves, although the reason for chromatic 
X-ray light-curve breaks lies in the tracking of the reverse-shock electron distribution and not 
in some fundamental/simple afterglow process.

 In this work, we model the broadband emission of afterglow 090510 by calculating the synchrotron 
emission from both shocks. 
 Kumar \& Barniol Duran (2010) and Corsi et al (2010) have identified the emission of this afterglow
at all wavelengths with synchrotron from the forward-shock. Analytically and numerically, we find that,
if that were true, then the X-ray emission after the 2 ks break would be too dim for the GeV flux
measured by Fermi-LAT at 100 s. To solve this incompatibility, the optical and GeV emissions must be
from different emission processes, and we propose that the latter is inverse-Compton scatterings.
The fast decay of the X-ray flux after the 2 ks break indicates that the X-ray is also inverse-Compton.
As shown by Panaitescu \& Kumar (2000), inverse-Compton scatterings in the forward-shock could be
the dominant emission process in the X-ray if the ambient medium is denser than about 100 
protons/${\rm cm^{-3}}$ but, so far, there is only one case (000926 -- Harrison et al 2001, Panaitescu 
\& Kumar 2002) where the X-ray afterglow light-curve and flux warranted its explanation with the synchrotron 
self-Compton model.

\section{Afterglow observations and forward-shock model details}
 
\subsection{Properties of afterglow 090510}

\hspace*{5mm}
 A successful model for the multiwavelength emission of GRB afterglow must account for the
following properties (using the notation $F_\nu \propto t^{-\alpha} \nu^{-\beta}$ for the 
power-law afterglow flux):
\vspace*{-3mm}
\begin{enumerate}
 \item an {\sl achromatic} light-curve break (at 2 ks after trigger), appearing at the same time in the
  optical and X-ray, with the optical flux slowly rising at a power-law index $\alpha_{o1} = -0.2 \pm 0.1$ 
%  (De Pasquale et al 2009 report a faster $t^{0.5 \pm 0.1}$ rise) 
  followed $\alpha_{o2} = 1.1 \pm 0.1$ after the break, and the X-ray flux decaying with
  $\alpha_{x1} = 0.74 \pm 0.03$ and $\alpha_{x2} = 2.2 \pm 0.1$,
 \item a GeV light-curve with a power-law decay index $\alpha_g = 1.4 \pm 0.1$ at earlier times (up to 100 s),
 \item the spectral slopes measured at all three frequencies: $\beta_o \simeq 1$ (at 100 s in Fig 1, 
     and at 1000 s in fig 1 of De Pasquale et al 2010), $\beta_x = 0.6 \pm 0.1$, and $\beta_g = 1.1 \pm 0.1$ 
     (De Pasquale et al 2010), 
 \item the relative optical/Xray/GeV fluxes.
\end{enumerate}
\vspace*{-2mm}

 At 100 s (which is the only epoch of simultaneous measurements), the relative fluxes can be quantified 
by the spectral slope $\beta_{xg} = 0.72 \pm 0.02$, thus $\beta_x \simeq \beta_{xg}$, indicating that 
the X-ray and GeV emissions could be from same spectral component. Then, $\beta_g > \beta_{xg}$ indicates
the existence of a spectral break not much below 1 GeV. 

% The optical--to--X-ray slope increases progressively from $\beta_{ox} = 0.13 \pm 0.06$ at 100 s to
%$\beta_{ox} = 0.83 \pm 0.03$ at 10 ks, owing to the optical flux decay being always slower than in the X-rays. 
 The simultaneity of the optical and X-ray light-curve breaks indicates that they have the same origin,
most likely related to the dynamics of the relativistic source. That the optical and X-ray flux decay
indices differ so much, both before and after the break, indicates that either
$(i)$ the optical and X-ray are from same spectral component {\sl and} there is a spectral break in 
      between them
$(ii)$ the emission at these two frequencies is from different spectral components.

\subsection{Possible afterglow models}

\hspace*{5mm}
 {\sl What not?}
 The temporal behaviour and spectrum of the GeV emission of afterglow 090510 provide strong reasons 
(Kumar and Barniol Duran 2010) to believe that it arises from the same mechanism as the X-ray emission. 
Thus, there is no reason to attribute the GeV emission to hadronic processes unless such processes 
are also believed to produce the lower-energy afterglow emission (X-ray and below). Additionally,
as shown by Asano et al (2009) and Razzaque (2010), synchrotron emission from protons and inverse-Compton
from secondary electron-positron pairs produced in photo hadronic interactions would require protons 
to acquire a total energy exceeding by 2--3 orders of magnitude the $\gamma$-ray output of GRB 090510.

 An origin of the delayed GeV emission of afterglow 090510 in late internal-shocks (Rees \& M\'esz\'aros 
1994) similar to those which may have produced the MeV prompt emission (as discussed by De Pasquale et 
al 2010) seems very unlikely, given that the spectra of the prompt MeV emission (described by a Band 
function peaking at $\sim 4$ MeV) and of the delayed GeV emission (a $F_\nu \propto \nu^{-1}$ power-law) 
are so different.
 
 {\sl Reverse-shock possible, forward-shock more likely}.
 An origin of the GeV emission in the reverse-shock crossing the ejecta that catch-up with the leading 
part of the outflow at 1--100 s is possible provided that the incoming ejecta drive a relativistic 
reverse-shock that accelerates electrons to a sufficiently high energy that their synchrotron and/or 
inverse-Compton emission can reach 1 GeV.
 The forward-shock model (M\'esz\'aros \& Rees 1997a) provides, in general, a more natural explanation for 
the long-lived, power-law decay of the afterglow flux. Thus, it seems more likely that the entire emission 
of GRB afterglow 090510 arises from the forward-shock, as was proposed by Kumar \& Barniol Duran (2010) and 
Ghirlanda et al (2010). In the numerical calculations of the blast-wave emission presented below, the
emission from both these shocks (reverse and forward) is taken into account.

\subsection{Numerical model for shock dynamics and emission}

\hspace*{5mm}
 The numerical model used to fit the multiwavelength emission of afterglow 090510 calculates
the dynamics (Lorentz factor $\Gamma$, mass) of the forward and reverse shocks and the synchrotron 
and inverse-Compton from them. The deceleration caused by the interaction with the ambient medium is 
calculated from that the swept-up fluid has a post-shock energy-to-mass ratio of $\Gamma$, thus
$\Gamma^2 M_{FS} = const$, which is corrected for the effects of radiative losses and outflow lateral 
expansion. For the latter, the kinetic-energy per solid-angle of the shock is approximated as uniform 
(same in all directions within the outflow opening). The thickness of the shocked gas is ignored in the
calculation of the photon arrival-time (i.e. the emitting fluid is approximated as a surface).

 The emission from relativistic electrons is calculated by assuming that they have a power-law 
distribution with energy, $dN/d\gamma \propto \gamma^{-p}$, above an electron energy $\gamma_*$ 
parametrized by the total electron energy being a fraction $\epsilon_e$ of the post-shock energy. 
Similarly, the magnetic field strength $B$ is parametrized by its fraction $\epsilon_B$ of the shock 
energy. 
 The electron distribution has a "cooling break" at an energy $\gamma_c m_e c^2$ where the radiative 
(synchrotron and inverse-Compton) timescale equals the dynamical one (over which the electron distribution
is replenished). Above $\gamma_c$, the effective electron distribution is 
$dN/d\gamma \propto \gamma^{-(p+1)}$ if $\gamma_* < \gamma_c$; otherwise $dN/d\gamma \propto \gamma^{-2}$ 
for $\gamma_c < \gamma < \gamma_*$ and $dN/d\gamma \propto \gamma^{-(p+1)}$ for $\gamma_* < \gamma$.

 The synchrotron spectrum has a peak at photon energy $\nu_p^{(sy)} \propto B\Gamma \min (\gamma_*^2,\gamma_c^2)$
and flux $F_p^{(sy)} \propto (dN_e/d\Omega) B \Gamma$, where $N_e$ is the number of electrons in the 
shock, when the outflow is sufficiently relativistic that $\Gamma > \theta_j^{-1}$, with $\theta_j$ the
jet half-angle; after $\Gamma < \theta_j^{-1}$, the synchrotron peak flux is $F_p^{(sy)} \propto N_e B \Gamma^3$.

 The inverse-Compton spectrum peaks at energy $\nu_p^{(ic)} = \min(\gamma_*^2 \nu_*, \gamma_c^2 \nu_c)$ 
and a peak flux $F_p^{(ic)} \sim \tau_e F_p^{(sy)}$, with a correction
factor that depends on the electron index $p$, $\tau_e$ being the shock's optical thickness.
Numerically, $F_p^{(ic)}$ is calculated from the Compton parameter $Y$ (ratio of the frequency-integrated 
inverse-Compton and synchrotron fluxes) and the synchrotron and inverse-Compton spectral breaks.
The $Y$ parameter depends on the cooling energy $\gamma_c$ if $p<3$ or if $\gamma_c < \gamma_*$ 
and $\gamma_c \propto (Y+1)^{-1}$. An analytical solutions for $Y$ and $\gamma_c$ can be found in the 
limits $\gamma_* \ll \gamma_c$ and $\gamma_c \ll \gamma_*$; however, these conditions are not always 
satisfied and the solutions are best calculated numerically. Further complications in the calculation 
of $\gamma_c$ and $Y$ arise in the case of a dense ambient medium, when the synchrotron emission may be 
self-absorbed up to a frequency higher than the peak of the synchrotron spectrum.

 The reverse-shock emission is calculated in the numerical model used to obtain the results presented 
below, and could be of importance at lower photon energies, particularly if energy is given to the 
forward-shock by means of some late ejecta, as the existence of the reverse-shock is guaranteed in that 
case. However, if the magnetic field and the electrons acquire the same respective fractions $\epsilon_e$ 
and $\epsilon_B$ of the fluid energy behind both shocks, we find that, for the numerical data fits 
presented below, the forward-shock emission is always brighter than the reverse-shock's at the photon 
energies of interest (optical, X-ray, and GeV). 
 
 The basic equations for the forward-shock dynamics and emission, used in the numerical model, are 
those presented by Panaitescu \& Kumar (2000, 2001), while those for the calculation of the Compton
parameter and electron cooling energy are given in Panaitescu \& M\'esz\'aros (1999).

\section{Forward-shock models for afterglow 090510} 

\subsection{Optical, X-ray, and GeV afterglow emissions are synchrotron -- uniform outflow}

\hspace*{5mm}
{\sl Host reddening is required}.
 An origin of the optical and X-ray emissions of this afterglow from the same radiation process is 
compatible with $\beta_o \simeq 1$ and $\beta_{ox} (100s) = 0.13 \pm 0.06$ only if the optical spectral
energy distribution (SED) is reddened by dust in the host galaxy (i.e. the intrinsic optical SED is harder).
 For a $A_\nu \propto \nu$ reddening curve, dust-extinction by 1.0--1.5 mag at a host-frame photon energy 
$\sim 3(z+1)$ eV would allow a $F_\nu \propto \nu^{1/3}$ intrinsic optical afterglow spectrum, 
and the optical and X-ray afterglow emissions could be from the same synchrotron 
spectral component, with the peak of the synchrotron spectrum located between optical and X-ray. 

{\sl Spectral break crossing optical at 2 ks is required}.
 If that spectral peak were the cooling frequency $\nu_c$, then the model expectation ($\alpha_1=-1/6$) 
for a homogeneous circumburst medium would match very well the observed optical flux rise, but the optical 
flux decay with $\alpha_{o2} = 1.1 \pm 0.1$ measured after the 2 ks break (associated with $\nu_c$ falling below 
the observing frequency) would be too fast for the expected $\alpha_2 = 1/4$. 
 Thus, the spectral peak located between optical and X-ray until the 2 ks light-curve break should be 
the characteristic synchrotron frequency $\nu_*$ at which radiate the $\gamma_*$-electrons.
Then, for $\nu_c > \nu_*$, the expected rise of the optical flux is $\alpha_1 = -1/2$ for a homogeneous
medium and $\alpha_1 = 0$ for a wind-like medium, both consistent with the UVOT light-curve until 2 ks.
The observed X-ray flux decay indicates that $h\nu_* (100s) < 1$ keV which, together with the expected 
evolution $\nu_* \propto t^{-3/2}$, implies that $\nu_*$ should fall below optical (3eV) no later than 
5 ks. This indicates that the optical light-curve break at 2 ks is due to the peak frequency $\nu_*$ 
falling below the optical.

 This is the model proposed by De Pasquale et al (2010) although, as they noticed, the model $F_\nu \propto
\nu^{1/3}$ optical spectrum is incompatible with the measured optical SED at 1 ks, which is much softer. 
While dust-reddening could solve that issue, there is no evidence for a substantial spectral softening by 
$\Delta \beta_o \simeq 1$ of the optical SED at 2 ks (see fig 3 of De Pasquale et al 2010), when $\nu_*$ 
decreases below the optical, although it could be argued that the large uncertainties of the UVOT 
measurements do not rule out such a spectral softening.

{\sl Coincidence: jet-break time occurs when $\nu_*$ crosses the optical.}
 While the decrease of the $\nu_*$ spectral peak from above to below optical frequencies can explain
the 2 ks optical light-curve break, a different factor must be producing the break in the X-ray
light-curve observed at about the same time. The fast decay of the post-break X-ray flux ($\alpha_{x2} = 
2.2 \pm 0.1$) suggests a jet-break occurring at that time, which requires an index $p = \alpha_{x2}$ 
of the power-law distribution of electrons with energy. Then, the slowest X-ray pre-break decay allowed
by the forward-shock model is $\alpha_1 = (3p-3)/4 = 0.9 \pm 0.1$ (obtained for a homogeneous medium
and X-ray below the cooling frequency), which is slightly faster than measured ($\alpha_{x1} = 0.74 \pm 0.03$).
The expected X-ray spectral slope ($\beta = (p-1)/2 = 0.60 \pm 0.05$) is consistent with observations.
The steeper decay of the GeV flux and its softer spectrum require that the cooling frequency is
below 100 MeV, for which the model expectations are 
$\alpha = (3p-2)/4 = 1.15 \pm 0.1$ (somewhat slower than measured until 100 s -- $\alpha_g = 1.4 \pm 0.1$), 
and $\beta = p/2 = 1.1 \pm 0.05$ (consistent with observations).

{\sl Above model is inconsistent with post-break optical flux decay}. 
 The jet model described above was discussed by De Pasquale et al (2010) and proposed by Kumar \& Barniol 
Duran (2010) and Corsi et al (2010). It is not a viable explanation for the X-ray light-curve break because 
it requires that the optical light-curve should exhibit a similar decay ($\alpha_{x2} = \alpha_{o2}$) after 
$\nu_*$ falls below the optical. 
In contrast, the optical flux decay is much slower than at X-rays ($\alpha_{x2} - \alpha_{o2} = 1.10 \pm 0.14$). 
Given that the slow $t^{-1}$ decay of the post-break optical flux is measured over a decade in time, 
the host galaxy contribution cannot be large enough to alter significantly the decay rate of the host-subtracted
afterglow flux, as was ruled out by De Pasquale et al (2010). 

{\sl Unnecessary numerical proof.}
 Fig 1 shows the best-fit to the optical, X-ray, and GeV emission obtained with the model above,
where the $\nu_*$ synchrotron spectrum peak crosses the optical at about the jet-break epoch.
As expected, the cooling frequency of the best fit model is between X-ray and GeV and the model X-ray
(GeV) light-curves decay slightly faster (slower) than observed.
Not surprisingly, the model optical and X-ray flux decays are the same after the 2 ks break and cannot
accommodate both post-break flux decays. The $\nu_*$ falling below the optical cannot occur 
significantly later than the optical light-curve peak shown in Fig 1 (left panel) because, as discussed 
above, $\nu_*$ must be below 1 keV at 100 s, to account for the X-ray flux decay measured at that time. 

\begin{figure*} 
\psfig{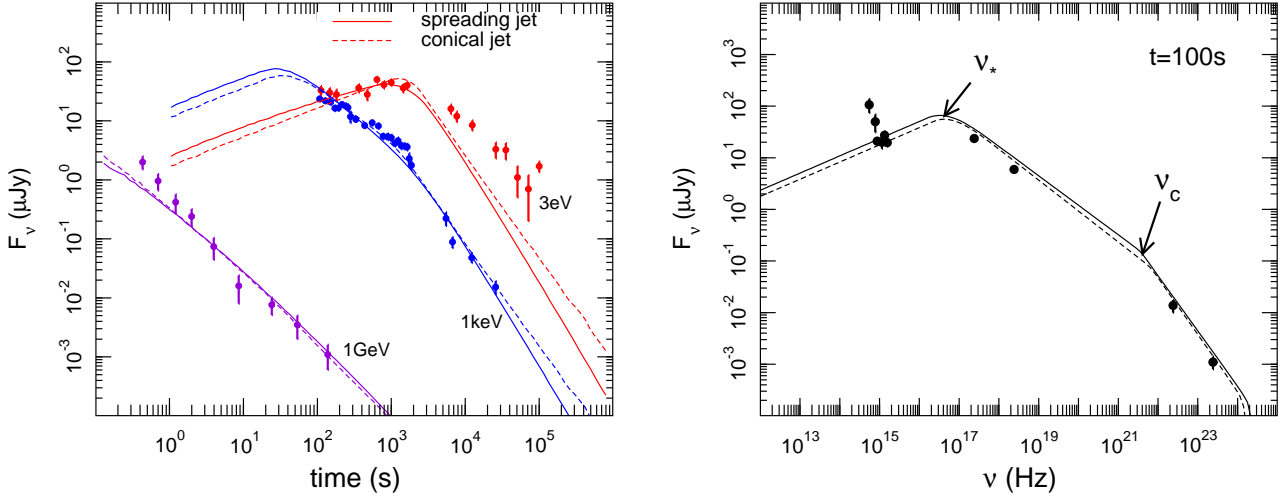} 
\caption{ Best fit to the optical, X-ray, and GeV emission of afterglow 090510, obtained with
  a numerical model that tracks the dynamical evolution of a relativistic jet decelerating by the
  interaction with a homogeneous ambient medium. The afterglow emission is forward-shock synchrotron 
  at all wavelengths. 
  The boundary of the jet becomes visible to the observer at 2 ks, when the jet 
  has decelerated to a Lorentz factor $\Gamma = \theta_j^{-1}$, with $\theta_j$ being the half-angle
  of the jet. The peak of the synchrotron spectrum falls below the optical at about the jet-break, 
  to account for the slow rise of the optical light-curve before 2 ks and its decay after that. 
  The cooling break of the synchrotron spectrum is between X-ray and GeV, to accommodate the faster 
  GeV light-curve decay and its softer spectrum. 
  Solid lines are for a jet spreading laterally, while dashed curves are for a conical jet (confined).
  Neither case can accommodate the post-break (slower) optical and (faster) X-ray flux decays. 
  Before the 2 ks break, the model optical spectrum is harder than measured, but that could be due 
  to dust-reddening in the host galaxy. 
  The best-fit outflow parameters are: ejecta initial kinetic energy-per-solid angle $E_0 = 10^{52.5}$
  erg/sr, ambient medium density $n = 10^{-4}\, {\rm cm}^{-3}$ (compatible with this short burst arising 
  from a NS-NS or NS-BH merger occurring outside the host galaxy), jet half-angle $\theta_j = 0.3 \deg$
  (unusually small when compared with the jets of other afterglows, but required by the early jet break
  time of only 2 ks, given the low ambient density), electron post-shock fractional energy $\epsilon_e
  = 1/4$, index of electron power-law distribution with energy $p=2.2$, magnetic fractional energy
  $\epsilon_B = 10^{-3.5}$. These parameters are similar to those inferred by Corsi et al (2010) from
  modelling the afterglow data and consistent with the parameter ranges inferred by Kumar \& Barniol Duran
 (2010) from matching the optical, X-ray, and GeV fluxes at one epoch.
 }
%\label{}
\end{figure*}

{\sl Caveats to the rescue?}
 In the calculation of model light-curves, the comoving-frame synchrotron spectrum is assumed to be 
power-law piecewise, with sharp breaks, and the only smoothing of those breaks is caused by the integration 
over the photon equal arrival-time surface. The other factor which stretches the spectral breaks -- 
the real shape of the electron distribution obtained by integrating over time the injected distribution -- 
is not included the numerical model, but we have verified that the assumed broken power-law electron 
distribution is a very good approximation for the 
electron distribution obtained by (1) injecting a power-law distribution with energy at every step in 
the propagation of the forward-shock and (2) following the electron adiabatic and radiative cooling.
To account for the slower decay of the optical light-curve of afterglow 090510 measured over a decade 
in time, the peak energy $\nu_*$ of the synchrotron spectrum would have to be spread over two decades in 
frequency (resulting from that $\nu_* \propto t^{-2}$ for a spreading jet), which seems very unlikely.
Thus, unless the peak of the real electron distribution is somehow stretched over a decade in electron 
energy, the optical emission will arise from the $\gamma > \gamma_*$ electrons soon after 2 ks, and the 
optical light-curve decay will quickly become the $F_o \propto t^{-p}$ expected for a collimated
spreading outflow (jet-break epoch is also at 2 ks).

\subsection{Optical, X-ray, and GeV afterglow emissions are synchrotron -- structured outflow}

 The basic idea of this model is that the synchrotron emission from one part of the outflow is dominant in 
the optical while another part of the outflow produces most of the X-ray emission.
 The important thing to understand about this model is that variations of the forward-shock parameters 
(energy, electron and magnetic energy parameters, external density) lead to afterglow spectra that 
{\sl do not cross} above the synchrotron spectral peak. 
 Consequently, if one part of the outflow is dominant at one frequency above that spectral peak then
it will be dominant at all frequencies above that peak. The only exception to this rule is when the
electron index $p < 2$ and for a varying magnetic field parameter $\epsilon_B$.
In this case, if the cooling frequency is between optical and X-ray, the optical and X-ray fluxes have an
opposite variation with $\epsilon_B$: $F_o \propto \epsilon_B^{(p+1)/4}$ and $F_x \propto \epsilon_B^{(p-2)/4}$, 
hence it would be possible for one part of the outflow with a higher $\epsilon_B$ to dominate the optical 
emission, while being dimmer in the X-rays than another part of the outflow of a lower $\epsilon_B$. 
However, this is not the case for GRB 090510 afterglow, because the GeV and the post-break X-ray flux decays 
require that $p>2$.

 Whatever is the angular structure of the outflow, to decouple the optical and X-ray light-curve decays 
by making one part of the outflow be dominant at only one frequency requires different optical-to-X-ray 
spectral slopes in the two outflow regions, otherwise one region would be dominant at both frequencies 
and the light-curve decay indices would be the same (if the cooling frequency $\nu_c$ is between optical 
and X-rays) or would differ by 1/4 (if $\nu_c$ is below optical or above X-rays). For observing frequencies 
that are in the same spectral region (above the peak frequency), as is the case for the optical and X-ray
fluxes of afterglow 090510 at ks after trigger, this is equivalent to requiring that the two outflows
have electron distributions with energy of different power-law indices $p$. 

 Current understanding of particle acceleration by relativistic shocks does not offer support
to the possibility that the electron distribution index is a function of the ultra-relativistic 
shock's Lorentz factor $\Gamma$. However, the X-ray afterglow spectral slopes measured by Swift have 
a distribution that is wide ($\Delta \beta_x \simeq 1/2$) and not bimodal (as expected for X-ray being 
either above or below cooling frequency), which indicates that relativistic shocks do not accelerate 
particles to a unique power-law distributions with energy for all afterglows. If relativistic shocks do 
so in different afterglows, then they could also yield electron distributions of different hardness in a 
structured outflow, where $\Gamma$ is not the same in all directions.

 However, even if a sufficient variation of the electron index $p$ with the shock Lorentz factor 
$\Gamma$ is possible and/or allowed by the decay indices and spectral slopes measured for a given
afterglow, obtaining decoupled light-curve decays at two different frequencies is not trivial because 
some fine-tuning of the other outflow parameters is required get one part of the outflow be dominant 
at only one frequency over a long period of time (about a decade for afterglow 090510), during 
which the shock decelerates and the afterglow spectrum softens. 

 For these reasons, a structured outflow, usually proposed as a top-hat (crown and brim) jet model
does not seem to be a viable explanation for GRB afterglows whose light-curves display very different 
power-law decays at two different frequencies, over the same period of time. 

 To reconcile the different decays of the optical and X-ray post-break light-curves, Corsi et al (2010) 
have proposed that the fast-decaying X-ray flux is produced by a narrower jet whose boundary becomes
"visible" (and lateral expansion begins to be important) after 2 ks, while the slowly-decaying optical flux  
arises from a wider outflow with a lower initial Lorentz factor that starts decelerating at 2 ks.
 We note that the electron index required by the slower decay of the optical flux from the wider outflow 
($p_{wide} = (4\alpha_{o2}+3)/3 \simeq 2.5$) is close to that required by the faster decay of the X-ray flux 
from the narrower, spreading jet ($p_{jet} = \alpha_{x2} \simeq 2.2$). 
Because $p_{wide} \simeq p_{jet}$, the optical and X-ray flux decays are not decoupled after the jet-break 
and the wider jet is dominant at both optical and X-ray frequencies after 2 ks, as can be seen in fig 2
of Corsi et al (2010).

\subsection{Optical, GeV, and only {\sl late} X-ray afterglow emissions are synchrotron (uniform outflow)}

\hspace*{5mm}
{\sl General model features.}
 We consider now the case of a wider outflow, for which the jet-break epoch is after the last measurement
of afterglow 090510 and the peak of the synchrotron spectrum crosses the optical at 2 ks. To account 
for the GeV flux decaying faster than the optical flux ($\alpha_g - \alpha_{o2} = 0.3 \pm 0.15$), the ambient 
medium must be homogeneous and the cooling frequency $\nu_c$ must be between optical and GeV (for which 
the model expectation is $\alpha_g - \alpha_{o2} = 1/4$). The index of the power-law distribution of electrons
with energy that accounts for the post-break optical flux decay is $p = (4\alpha_{o2}+3)/3= 2.47 \pm 0.13$. 
Then the X-ray spectrum will have a slope $\beta_x = (p-1)/2 = 0.74 \pm 0.06$ consistent with the observations
if $\nu_c$ is above X-ray (GeV spectral slope, $\beta_g = p/2 = 1.24 \pm 0.06$, is also consistent with the 
data). Thus, the observed flux decay indices and spectral slopes of afterglow 090510 could be explained 
with synchrotron forward-shock emission provided that the ambient medium is homogeneous and the cooling 
frequency is between X-ray and GeV.

{\sl Central-engine contribution to early X-ray afterglow}.
 The obvious problem with this model is that it cannot explain the X-ray break at 2 ks. 
One possible solution is that the early X-ray afterglow flux is not from the forward-shock, as for
other afterglows whose early X-ray light-curves exhibit a "plateau" which ends with a chromatic break, 
not seen in the optical light-curve. The lack of an evolution of the X-ray spectrum at the time of the
X-ray light-curve break rules out the passage of a spectral break through the X-ray band as the reason 
for that X-ray light-curve chromatic break, and suggests that the X-ray plateau arises from the same 
mechanism ("central engine") that produced the prompt emission, but operating for longer times at X-ray 
energies. However, the simultaneity of the breaks in the optical and X-ray light-curves of afterglow 
090510 strongly suggests that both emissions have the same origin and (unless the optical emission is
also attributed to the central engine) that the X-ray flux is not produced by the central engine.

{\sl Model failure}.
 Ignoring the above-discussed implication of the simultaneity of the optical and X-ray light-curve
breaks seen for the afterglow 090510, we consider the possibility that its early X-ray emission is 
not from the forward-shock, and model only its optical, GeV, and the X-ray emission after the plateau, 
as the last four XRT measurements at 5--25 ks display a power-law decay of index $\alpha_x = 1.43 \pm 0.22$ 
that is compatible with the optical and GeV flux decays. The result is shown in Fig 2 (solid lines): 
the best-fit underestimates the GeV flux and, owing to the very low medium density, the cooling break 
is above GeV (but that is the lesser failure). If the errors of the GeV measurements are reduced to 
force the numerical model to accommodate the GeV light-curve, then the model X-ray flux exceeds 
observations (dashed lines).

\begin{figure*}
\psfig{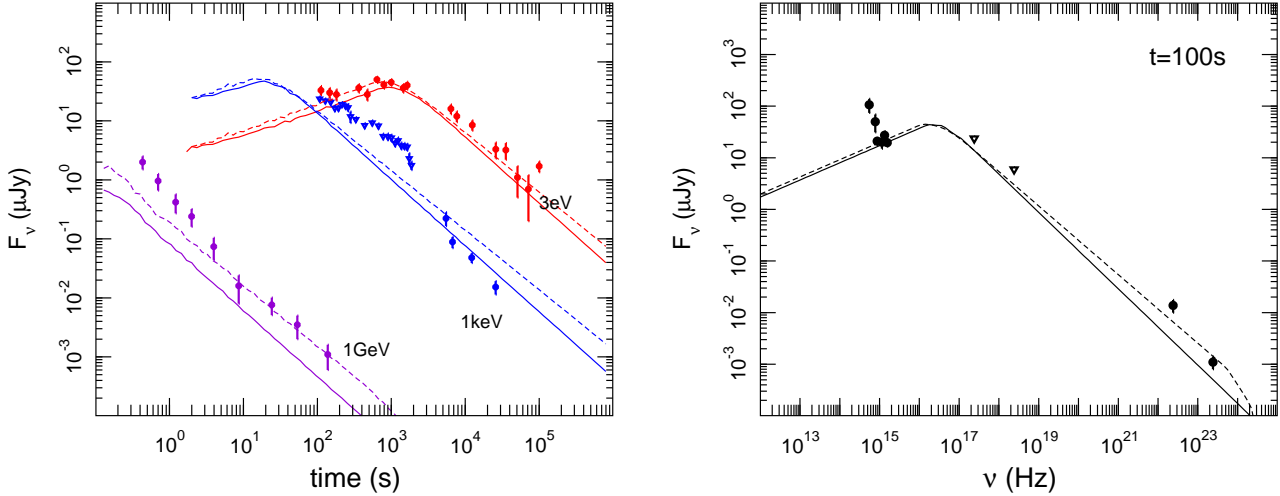}
\caption{ Best fit to the optical, late X-ray, and GeV emission of afterglow 090510, obtained with
 forward-shock synchrotron model for a wide outflow (jet-break after last measurement) and the peak frequency 
 of the synchrotron spectrum crossing the optical at 2 ks. Only the X-ray data after 2 ks is included in
 the fit, while the X-ray measurements prior to 2 ks are attributed to the central engine, and represent
 only upper limits for the forward-shock emission.
 Solid lines show the best-fit for the actual errors of the measured GeV fluxes; the model cannot account 
 for those measurements. 
 Parameters are: $E_0 = 10^{52.7}$ erg/sr,  $n = 10^{-6.5}\, {\rm cm}^{-3}$, $\epsilon_e = 0.07$, $p=2.5$, 
 $\epsilon_B = 10^{-4}$.
 Dashed lines are for the model obtained when the GeV measurement errors have been reduced by a factor 5, 
 to force the model to accommodate the GeV measurements; the model fluxes overestimate the X-ray measurements
 at 5--25 ks.
 }
%\label{}
\end{figure*}

{\sl Reason for failure: mismatch of X-ray and GeV fluxes}.
 This indicates that there is a simple problem with the synchrotron forward-shock emission model for the
afterglow 090510: the X-ray flux after 2 ks is too small for the GeV flux measured until 100 s, which can 
be understood in the following way.
The back-extrapolation of the X-ray flux after 2 ks to the epoch of the last GeV measurement implies that 
the forward-shock synchrotron X-ray flux at 100 s (overshined by the mechanism dominating the early X-ray
afterglow) was $F_{1 keV}(100s) = F_{1 keV}(10 ks) (10ks/100s)^{\alpha_x}$.
This X-ray flux can be extrapolated to GeV using the X-ray spectral slope $\beta_x$. 
The cooling frequency $\nu_c$ must be between optical and GeV (and the ambient medium must be homogeneous) 
to account for the decay of the GeV flux being faster than in the optical (after the 2 ks break). 
Then, there are two possibilities: (1) $\nu_c$ is between X-ray and GeV and (2) $\nu_c$ is between optical
and X-ray. 
In the former case (1), the extrapolation of the X-ray spectrum to GeV using the X-ray spectral slope 
yields an upper limit on the GeV flux: $F_{1GeV}(100s) < F_{1keV}(100s) (1keV/1GeV)^{\beta_x}$.
For a homogeneous medium and cooling above X-ray, $\alpha_x = \alpha_{o2} = 1.5 \beta_x$, 
we arrive at $F_{1GeV}(100s) < 10^{-2\alpha_{o2}} F_{1keV}(10 ks)$; the optical flux decay index $\alpha_{o2}$ 
was used instead of the X-ray flux decay index $\alpha_x$ because the former is better constrained by 
observations. Thus, for the measured $F_{1 keV}(10ks) \simeq 65$ nJy, the expected upper limit on the GeV 
flux is $F_{1GeV}(100s) < 0.4 \pm 0.2$ nJy, with the uncertainty calculated from the optical flux decay 
index $\alpha_{o2}$. This upper limit is a factor $\sim 4$ lower than the GeV flux measured at 100 s, of 
about $1.7 \pm 0.5$ nJy, thus there is an excess of GeV flux. 
In the latter case (2), the extrapolation of the X-ray spectrum to GeV is an exact estimation of the GeV
flux. Using $\alpha_x = \alpha_{o2} + 1/4$ and $\alpha_x = (3\beta_x-1)/2$, we arrive at
$F_{1GeV}(100s) = 10^{-2\alpha_{o2}-2.5} F_{1keV}(10 ks)$, which is smaller by a factor 300 than the above
upper limit and well below the GeV flux measured at 100 s.

{\sl Possible solution}.
 This incompatibility between the X-ray and GeV fluxes of afterglow 090510 indicates that the optical, 
X-ray, and GeV emissions are not from the same spectral component. If optical and X-ray are not from the
same emission process but X-ray and GeV are, then the indices $\alpha_x$ and $\beta_x$ could be larger 
than implied by $\alpha_{o2}$ and the extrapolated X-ray flux could also be larger and compatible with 
the measured GeV flux. Alternatively, if the optical and X-ray emissions are from same component, then
the X-ray flux extrapolated to GeV being smaller than measured indicates that the GeV emission arises 
from another spectral component.

 Either of the above possibilities takes us to a model where the lower-frequency (optical and, maybe, 
X-ray) afterglow emission is synchrotron and the higher-frequency (GeV and, possibly, X-ray) emission 
is inverse-Compton. For the inverse-Compton emission to be brighter at GeV than the synchrotron, 
the external medium density (which sets the optical thickness to electron scattering in the 
forward-shock) must be much denser than for the models shown in Figs 1 and 2.

\subsection{Optical is synchrotron, GeV is inverse-Compton, and X-ray is ? (uniform outflow)}

\hspace*{5mm}
 The best-fit obtained with a dense ambient medium and inverse-Compton emission for the high
frequency afterglow is shown in Fig 3. 
 For either type of medium, homogeneous or wind-like, the best-fit has the inverse-Compton brighter
than synchrotron in the X-ray. 
 Given that the angular structure of the outflow can account
for the very different rates of decay of the X-ray and optical light-curves after the 2 ks break,
we have considered another possibility for altering the outflow dynamics and obtaining an achromatic
light-curve break: cessation of energy injection in the blast-wave 
In this model, energy is added to the blast-wave when some later-released ejecta arrive at it,
the flux decay prior to the break (i.e. end of energy injection) being determined by the rate at 
which energy is added while, after the break, the decays are those expected for an adiabatic, 
"spherical" outflow (i.e. a jet sufficiently wide that the emission from its boundary is not 
relativistically beamed toward the observer: $\theta_{jet} > \Gamma^{-1}$).

\begin{figure*}
\psfig{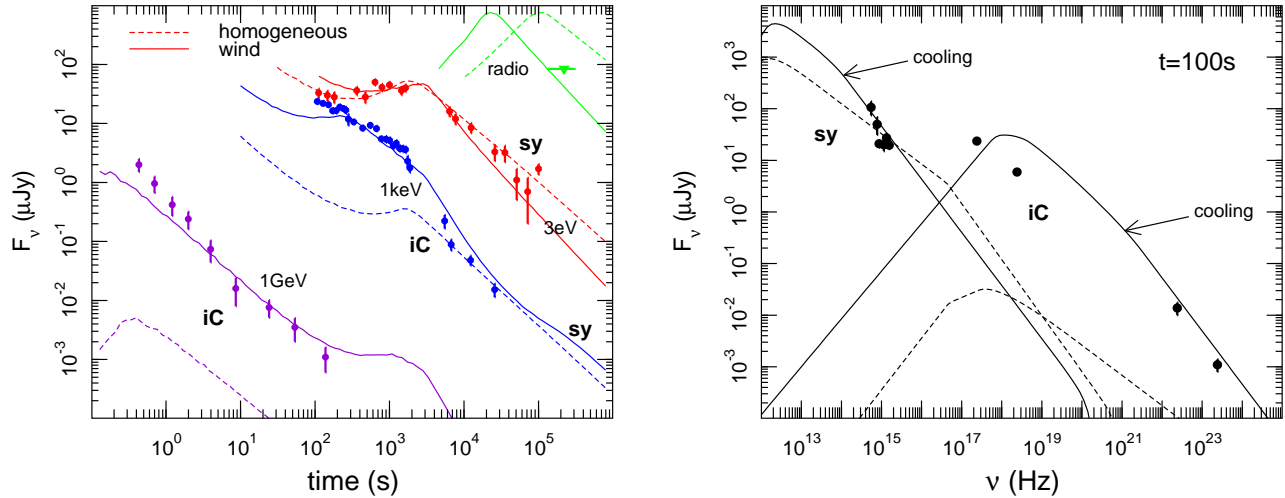}
\caption{ Best fit to the optical, X-ray, and GeV emissions of afterglow 090510, obtained with
 the synchrotron self-Compton (SsC) forward-shock model, for a high external medium density. 
 The higher frequency emission (GeV, in the least) is inverse-Compton, while the lower frequency emission 
 (radio and optical) is synchrotron. The X-ray flux is dominated by inverse-Compton scatterings for either 
 type of medium (dashed lines = homogeneous, solid lines = wind; spectral components are shown in the right 
 panel, with same line coding).
 % with the exception of the slower decay exhibited after 50 ks by the light-curve for a wind medium, 
 % when synchrotron emission becomes brighter than the faster decaying inverse-Compton flux.
 The achromatic light-curve breaks at 2 ks and the slower pre-break decays are modeled with energy injection
 in the forward-shock, which is terminated at 2 ks. For a homogeneous medium, the SsC model fails to account 
 for the pre-break X-ray and GeV emission. For a wind medium, the SsC model accounts for the major light-curve 
 behaviours at all three frequencies and yields a radio flux (at 8 GHz) below the $3\sigma$ upper limit 
 reported by Frail (2009).
 For the latter case, parameters are: initial ejecta kinetic energy $E_0 = 10^{52.6}$ erg/sr, 
 kinetic energy of the incoming ejecta $E_{inj} = 10^{53.8}$ erg/sr, energy is added to the shock up to 2 ks 
 according to a $E \propto t^{1.2}$ law ($t$ is observer time), $\epsilon_e = 0.15$, $p=2.1$, $\epsilon_B = 
 10^{-10}$, and the ambient medium density has a $n \propto r^{-2}$ radial stratification with the 
 normalization corresponding to a GRB progenitor mass-loss rate--to--terminal wind-speed ratio of 
 $2.10^{-4} (M_{sun}/yr)/(1000 km/s)$.
 }
%\label{}
\end{figure*}

%\vspace*{-5mm}
\subsubsection{Homogeneous medium}

\hspace*{5mm} 
 This model clearly fails, yielding (pre-break) X-ray 
and GeV fluxes that are well below those measured. For higher external densities, around $n \simeq 
10^6 {\rm cm}^{-3}$, the model inverse-Compton emission matches the measured X-ray and GeV fluxes 
but the fit to the optical light-curve becomes poor because the ever-softening inverse-Compton spectrum 
becomes dominant in the optical after the break and exceeds the measured optical flux. 

 The overproduction of optical flux in the synchrotron self-Compton (SsC) model, for a homogeneous medium,
and for inverse-Compton matching the X-ray measurements of afterglow 090510, is caused by that 
the peak flux of the inverse-Compton spectrum increases as $F_p^{(ic)} \propto E^{5/4} t^{1/4}$. 
Then, if the inverse-Compton flux at at 100 s and 1 keV accounts for the measured  $F_x \sim 20\,\mu$Jy, 
the inverse-Compton peak flux will be $F_p^{(ic)} > 20\,\mu Jy \times (2ks/100s)^{(5e+1)/4}$ at the 
break-time of 2 ks (when energy injection ends), where the energy injection follows a $E \propto t^e$ law.
The exponent $e$ can be determined from the break in the optical light-curve that cessation of
energy injection should produce: $\alpha_{o2} - \alpha_{o1} = (p+3)e/4$, with the electron index $p$
constrained from the post-break optical flux decay: $\alpha_{o2} =(3p-3)/4$. Thus $p \simeq 2.5$, 
$e \simeq 1.0$, and  $F_p^{(ic)} (2ks) > 1.8$ mJy. At that time (2ks), the peak energy of the inverse-Compton
spectrum is around the optical, yielding an optical flux that will exceed by much the observed $\sim 50\,\mu$Jy 
flux. 

 This means that if the inverse-Compton emission matched the pre-break X-ray flux, it would over-predict 
the post-break optical emission. Most of that overproduction of optical flux arises from the substantial
energy increase ($E \propto t$ from 100 s to 2 ks), by more than a factor 10, required to account for the
pre-break optical flux rise and X-ray plateau. 

 A second reason for which the SsC model with a homogeneous medium would fail is that it cannot accommodate
the X-ray flux decay after the 2 ks break. The steepest decay that this model can yield, $\alpha_{x2} = 
(9p-10)/8$, is only $\alpha_{x2} = 1.6$ (for the electron index $p \simeq 2.5$ required by the post-break
optical flux decay), which is too slow compared to the observed X-ray flux decay ($\alpha_{x2}=2.2$).

%\vspace*{-5mm}
\subsubsection{Wind-like medium}
\label{windmodel}

\hspace*{5mm} 
 A wind-like ambient medium provides a much better fit to the multiwavelength data and displays all
the required features, although it has some shortcomings: post-break optical flux decay is slightly 
faster than measured, GeV flux decay is slightly slower than observed, post-break X-ray flux is 
over-predicted, and the earliest optical, X-ray, and GeV measurements are not well matched.
The reasons for some of these model successes and limitations are described below. In contrast with the
case of a homogeneous medium, the inverse-Compton emission can match the pre-break X-ray flux without
overproducing optical flux after the break (when the decreasing inverse-Compton peak energy reaches
optical frequencies) because the inverse-Compton peak flux decreases with time as $F_p^{(ic)} 
\propto E^0 t^{-1}$ (independent of the shock's energy).

 After 1 ks, the synchrotron cooling frequency (which evolves $\nu_c \propto t^{1/2}$) rises above optical,
% if the Compton parameter $Y < 1$ but, here, $Y \gg 1$ and its decrease yields an increase of
% $\nu_c \propto Y^{-2}$ faster than $\nu_c \propto t^{1/2}$), 
thus the post-break optical flux decay $\alpha_{o2} = (3p-1)/4$ requires an electron index $p = 1.8 \pm 0.18$. 
 The X-ray domain is below the cooling frequency of the inverse-Compton spectrum 
(see Fig 3), hence the post-break X-ray flux decay index is $\alpha_{x2}= p$, observation requiring thus 
that $p = 2.2 \pm 0.1$, for which the X-ray spectral slope will be $\beta_x = (p-1)/2 = 0.6$ (as observed). 
 Before 100 s, the GeV range is above the cooling frequency of the inverse-Compton spectrum. 
As the shock energy is increased by a factor $E_{inj}/E_0 = 15$ according to $E \propto t^e$ with $e = 1.2$ 
until 2 ks (after which $e=0$), the energy injected prior to 100 s is less than the initial energy $E_0$ of 
the blast-wave. When there is no significant energy injection (and if radiative losses are also negligible),
the expected GeV flux decay index is $\alpha_g = p-1$, hence the data require that $p = 2.4 \pm 0.1$, 
for which the GeV spectral slope will be $\beta_g = p/2 = 1.2$ (as measured). 

 Therefore, in the SsC model with a wind-like medium, the GeV measurements at 1-100 s require a slightly
harder electron index than the optical data after 2 ks. Such a hardening of the post-shock electron
distribution is not included in the numerical model, whose best-fit electron index $p = 2.1$ is close 
to that required by the X-ray light-curve and spectrum and between the values required by the optical 
and GeV decay indices ($p=1.8$ and $p=2.4$, respectively). For that reason, the model optical 
flux decay is faster than observed while the model GeV flux decays slower than measured. 

 As these results are for an adiabatic blast-wave, it may seem possible that a highly radiative
forward-shock could account for the steeper decay of the early GeV flux (as proposed by Ghisellini 2010). 
In that case, the decay index 
of the inverse-Compton flux below the cooling frequency is $\alpha_g = 7p/6-1$, hence the measured GeV 
flux decay requires $p = 2.1 \pm 0.1$, which is consistent with the value required by the post-break
X-ray decay (and is also the best-fit value). Radiative losses are taken into account in the numerical 
calculation, however the resulting GeV flux decay is not as fast as observed because radiative losses 
are negligible, owing to the electron fractional energy being substantially below unity (i.e. even if 
electrons were fast cooling, they would release a quasi-negligible fraction of the shock-energy over 
one dynamical timescale; furthermore, the electrons containing most of the post-shock energy are slowly
cooling, as shown by the spectra of Fig 3).

 For the best-fit values $p=2.1$ and $e=1.2$, the model optical flux decay during energy injection should 
have a power-law index $\alpha_{o1} = [(3p-2)-(p+2)e]/4 = -0.15$ (compatible with the observed rise), 
while the expected X-ray flux decay index $\alpha_{x1} = p-0.5(p-1)e = 1.4$ is larger than measured 
(see Fig 2).

 {\bf IF} the SsC model with a wind medium is the correct explanation for the afterglow 090510 
multiwavelength emission, based on that it can account for all major features of this afterglow, 
and despite that the model light-curves shown in Fig 3 are not a perfect fit to the data, then the 
following model features stand out: \\
 (1) As a jet-break is not observed in the optical light-curve during the first day after trigger, 
the jet half-angle must be larger than $\theta_j \simeq 4.0 \deg$ (for the best-fit wind density and 
shock energy), which implies a jet energy larger than $E_{jet} = 10^{52}$ erg. This is 10 times higher 
than inferred for most other afterglows but does not exceed what the core-collapse of a massive star 
could release in a relativistic outflow, \\
 (2) The best-fit wind density is 3 times larger than the highest measured for Galactic Wolf-Rayet stars
by Nugis \& Lamers (2000); the quality of the best-fit is substantially poorer for a wind mass-loss--to--speed
ratio less than 3/4 of that of the best-fit shown in Fig 3 (for which the model GeV fluxes underestimate
all measurements). However, the medium where the afterglow emission is produced (less than 1/4 pc at 1 day) 
is shaped by the wind properties (mass-loss rate and terminal velocity) of the Wolf-Rayet star during its 
last 1\% life prior to the core collapse, a phase when the actual wind properties are not known.

 The low value of the magnetic field parameter $\epsilon_B = 10^{-10}$ of the best-fit corresponds to a 
post-shock magnetic field strength that can be accounted for by the shock-compression of an upstream magnetic 
field of about 0.1 mG, which corresponds to a fraction $\epsilon_B/\beta_w^2 \sim 10^{-4}$ of the wind 
kinetic energy, for the magnetic field parameter $\epsilon_B$ of the best fit and for a wind speed $\beta_w c 
= 1000$ km/s, thus it is plausible that the shock-generation of the magnetic field is not required to account 
for GRB 090510 afterglow emission. 
 
 The most important result obtained by modelling the multiwavelength emission of afterglow 090510 is 
that the ambient medium radial stratification \& density and the outflow energy are incompatible with a 
merger of two compact stars as the progenitor of this short GRB. Instead, those parameters point toward an 
origin from a massive star.

%\vspace*{-5mm}
\subsubsection{Pair-loading and acceleration of ambient medium by the GRB emission}

 We note that the model employed here to calculate the afterglow emission ignores that, up to a certain
radius, the ambient medium is lepton-enriched and accelerated by the pairs formed when GRB photons interact 
with the burst photons scattered by electrons in the ambient medium. The acceleration is due to both the 
scattering of GRB photons and the formation of pairs. The resulting "pair-wind" left behind by the GRB 
photon front is ahead of the GRB ejecta, delays the dissipation of their kinetic energy, and changes 
the dynamics of the afterglow blast-wave at early times.

 Beloborodov (2002) has shown that the number of pairs $n_\pm$ per ambient electron created through 
this process increases exponentially with distance $x$ measured inward from the leading edge of GRB photon
shell, $n_\pm \propto e^{x/x_\pm}$, where the loading length-scale is $x_\pm \sim (24 \div 33) \lambda$,
$\lambda = (m_e c^3)/(F_\gamma \sigma_T)$ being the electron free-path. Here $F_\gamma = E_\gamma/
(4\pi R^2 t_\gamma)$ is the flux of GRB photons above $m_e c^2$ at radius $R$, $E_\gamma$ and $t_\gamma$ 
being the burst energy release above $m_e c^2$ and the burst duration, respectively. 
Beloborodov (2002) has also shown that the acceleration length-scale, defined by the pair-enriched medium
acquiring a Lorentz factor $\gamma = 2$, is $x_a \simeq 200 \lambda$, both results for $x_\pm$ and
$x_a$ holding for a peak energy of the burst spectrum of $\sim m_e c^2$, which is typical for most GRBs 
(but not for GRB 090510).

 Pair-enrichment and acceleration of the ambient medium by the GRB photons are important up to a radius 
$R_\pm$ (or $R_a$) where the length-scale $x_\pm$ (or $x_a$) is equal the radial width of the GRB photon 
pulse, $\Delta = c t_\gamma$. The term $\lambda /\Delta = (4\pi m_e c^2/\sigma_T) (R^2/E_\gamma) = 0.015\,
R_{16}^2/E_{\gamma,53}$ shows that $R_\pm$ and $R_a$ depend only on the burst energy. For GRB 090510, 
$E_\gamma = 10^{52.6}$ erg, thus $R_{acc} = 10^{15.5}$ cm and $R_\pm \simeq 10^{16}$ cm.
 The acceleration of the ambient medium by scattered burst photons delays the onset of the blast-wave 
deceleration until $R_{acc}$, leading to a sharp rise in the afterglow emission, as shown by the 
bolometric light-curve of fig 5 of Beloborodov (2002). From $R_{acc}$ to $R_\pm$, the ambient medium
is nearly static, but pair-enriched. After $R_\pm$, the circumburst medium is static and pair-free,
with the afterglow emission being as calculated in the previous section. 
 
 For the wind-medium model discussed in \S\ref{windmodel} and shown in Figure 3, the blast-wave energy 
is initially $E_0 = 10^{52.6}$ erg, increasing with observer time as $E \propto t^{1.2}$ until $t_i = 2$ ks,
to a total energy $E_i = 10^{53.8}$ erg. Therefore, the injected energy is negligible until $t_0 = t_i 
(E_0/E_i)^{0.8} = 0.1\, t_i = 200$ s; until $t_0$ the blast-wave energy is the initial $E_0$ and its
radius increases as $R_{aglow} = 10^{15.3} (t/1\,s)^{1/2}$ cm. Here, we have used the arrival-time of 
photons emitted from the edge of the visible region (i.e. moving at angle $\theta = \Gamma^{-1}$ relative 
to the direction toward the observer). Thus, for the model discussed in \S\ref{windmodel}, $R_{aglow} (t_0) 
= 10^{15.8}$ cm when energy injection becomes dynamically important, and $R_{acc} \siml R_{aglow} (t_0) 
\siml R_\pm$, therefore the pair-loading and acceleration of the ambient medium by burst photons modifies 
the model afterglow light-curves shown in Fig 3 by suppressing the forward-shock emission until about $t_0 =
200$ s, after which the afterglow emission is as shown in Fig 3. That effect is only mildly inconsistent 
with the optical and X-ray fluxes of GRB afterglow 090510, which have been measured starting from 100 s, 
but is incompatible with the afterglow GeV flux, which decays from 0.5 s (first measurement) until about 
200 s (last measurement). 

 Thus, if the acceleration of the ambient medium by burst photons operates as envisaged by Beloborodov (2002), 
then the GeV emission of GRB afterglow 090510 measured at 0.5--200 s cannot arise from the blast-wave model 
of \S\ref{windmodel}.
 That is a general fact pertaining to afterglows produced by relativistic blast-waves interacting with 
Wolf-Rayet winds, for which the afterglow radius, $R = 1.3 \times 10^{15} (E_{a,53} t)^{1/2}$ cm 
(corresponding to a wind with a $10^{-5} M_{sun}/yr$ mass-loss rate and a 1000 km/s terminal velocity,
and $E_a$ being the blast-wave kinetic energy), is below the acceleration-radius, $R_\pm = 6 \times 10^{15} 
E_{\gamma,53}^{1/2}$ cm, until an observer time $t_a = 20\, (z+1) (E_\gamma/E_a)$ s.
As the pair-wind delays the dissipation of the blast-wave energy until after $t_a$, there is no significant
emission produced by the blast-wave prior to $t_a$.
 
 However, there are ways in which the effect/importance of the pair-wind may be overestimated:

{\sl 1. GRB photons of energy $m_e c^2$}.
 The intrinsic spectral peak-energy of GRB 090510 is around 6 MeV, which is significantly higher 
than the $\epsilon_\gamma \simeq m_e c^2$ assumed by Beloborodov (2002). 
In this case, the scattering of most GRB photons by the static, ambient electrons (just entering 
the photon front) occurs in the Klein-Nishina regime. Denoting by $y=\epsilon_\gamma/m_e c^2 \gg 1$, 
the electron-scattering cross-section is $\sigma_e \sim \sigma_T/y$, the electron takes most of the 
photon's energy ($\epsilon_e = (y/y+1)\epsilon_\gamma)$, while the scattered photon retains a small 
fraction of its initial energy ($\epsilon'_\gamma = \epsilon_\gamma/y$) and is scattered at a small 
angle ($d\sigma/d\Omega$ has an angular-scale  $\theta_{sc} = \sqrt{2/y}$), thus the threshold energy 
of a target photon to form pairs with the scattered photon is $\epsilon_{th} = 2\, m_e c^2/
(1-\cos \theta_{sc}) = 2 \epsilon_\gamma$.
 Compared to the case when $\epsilon_\gamma \simeq m_e c^2$, the reduced electron-scattering 
cross-section and the higher photon threshold-energy for $y \gg 1$ imply an increased pair-loading 
radius $R_\pm$; the lengthscales for acceleration by pair-formation and by photon scattering are 
also larger (for the latter, we note that $\sigma_e \epsilon_e \propto \ln (2y)/y$ is decreasing
with increasing photon energy $y$), hence $R_{acc}$ is increased.
 An accurate calculation of $R_{acc}$ and $R_\pm$ for $\epsilon_\gamma \gg m_e c^2$ is now
conveniently declared to be "beyond the scope of this paper". Instead, a crude estimate of $R_{acc}$ 
and $R_\pm$ can be obtained by using the burst output in the $(1-2)\,m_e c^2$ range, for which the 
electron scattering occurs at the limit between Thomson and Klein-Nishina regimes, and which 
Beloborodov (2002) found to be driving most of the pair-loading and acceleration. For GRB 090510, 
$E_\gamma (0.5-1 MeV) \simeq 0.04 E_\gamma (1-20 MeV)$, where $E_\gamma (1-20 MeV) = 3.6 \times 
10^{53}$ erg is the burst energy inferred from the measured fluence (Guiriec et al 2009).
A reduction by a factor 25 in $E_\gamma$ implies a reduction by a factor 5 in $R_{acc}$ because
$R_{acc} \propto E_\gamma^{1/2}$. Together with that $R_{aglow} \propto t^{1/2}$ (valid until 200 s,
when energy injection becomes dominant), it implies a reduction by a factor 25, to $t'_a = t_a/25 = 8$ s,
of the time until when the medium acceleration by the GRB photons is relevant. Still, the GeV emission 
of GRB afterglow 090510 is detected at even earlier times.

{\sl 2. Internal interactions in the pair-wind}.
 The interaction of the ultra-relativistic pair-wind produced by GRB photons at smaller radii with the
medium crossed by the GRB photon front at larger radii can reduce the acceleration and enrichment 
of the pair-wind produced at larger radii if that interaction happens within the GRB photon pulse.
That is so because the leptons swept-up by the earlier produced pair-wind move radially outward and
$i)$ will retain a smaller fraction of the seed photon energy, which reduces the momentum deposition 
in the wind, and 
$ii)$ will scatter the incoming photons, streaming outward, at a lower rate and at a smaller deflection 
angle, both reducing the rate of pair-formation.
 
 Together with the dilution of the photon front owing to its radial expansion, these were taken into 
account in the numerical model of Kumar \& Panaitescu (2004), who found (see their fig 1) that, for 
a GRB pulse of $10^{53}$ erg, spectrum peak energy of $m_e c^2$, and spectral energy distribution 
$F_\epsilon \propto \epsilon^{-1.5}$ above the peak (i.e. the parameters used by Beloborodov 2002), 
the acceleration radius is $R_{acc} = 10^{15}$ cm, while the pair-loading radius is $R_\pm = 10^{16}$ cm. 
Thus, when internal interactions in the pair-wind are taken into account, the resulting $R_{acc}$ is 
about 3 times smaller. That corresponds to a reduction by a factor 9 in the time up to which the medium 
acceleration is important.
If that reduction is applied to the shorter $R_{acc}$ estimated above from the GRB 090510 output in the 
$(1-2)\, m_e c^2$  range, then the onset of the afterglow and blast-wave deceleration for GRB 090510 should 
be seen at $t''_a = t'_a/9 \simeq 1$ s after trigger. Figure 1 of Ghirlanda et al (2010) shows that the
GeV light-curve of this afterglow has a peak at an even earlier time, at 0.3 s after trigger.
 
{\sl 3. Anisotropic GRB ejecta}.
 From the above, it may be concluded that, even after taking possible factors that reduce the pair-wind
acceleration radius, the GeV emission of GRB afterglow 090510 displays a peak that occurs too early to
be explained by the onset of the blast-wave deceleration. However, there is another reason for which the 
effect of the pair-wind on the blast-wave dissipation and afterglow emission may be less important. That 
reason has to do with that, during the burst phase, the observer may receive emission from a region of the 
GRB ejecta which is substantially narrower (in angle) than during the afterglow phase. That GRB pulses are
less time-asymmetric (fast rise, slow decay -- Norris et al 1996) than if the entire visible region of 
angular extent $\theta_\gamma = \Gamma_\gamma^{-1}$ (as measured from the fireball origin) were radiating
(Fenimore et al 1996), where $\Gamma_\gamma$ is the Lorentz factor of the GRB ejecta during the prompt 
emission phase, suggests that the burst emission arises from hot-spots of angular extent much less that 
$\theta_\gamma$. The pair-wind affects the deceleration of the ejecta only in these hot-spots, but does 
not accelerate the ambient medium ahead of the ejecta around the hot-spots, hence the afterglow emission 
from the region of angular extent $\theta_{aglow} = \Gamma_{aglow}^{-1}$ visible during the afterglow 
phase (where $\Gamma_{aglow} < \Gamma_\gamma$ is the afterglow Lorentz factor) may be unaffected by the 
pair-wind. 

% Concluding, the effect of the pair-wind created by the burst photons on the blast-wave deceleration and
%afterglow emission was ignored in the models presented in previous sections, however the GeV emission
%of GRB afterglow 090510 (whose power-law decay indicates an origin in the forward-shock) shows that that 
%effect may be present only up to most 0.3 s (when the GeV flux peaks).

%\vspace*{-3mm}
%\section{Some conclusions, all set in stone}
\section{Conclusions}

\hspace*{5mm}{\sl Uniform jet}.
 The salient properties of afterglow 090510 -- a slow rise of the optical flux, an {\sl achromatic} 
light-curve break at 2 ks, followed by a flux decay that is substantially slower in the optical than at 
X-rays -- cannot be explained with the traditional jet model (synchrotron emission from the forward-shock 
driven by a collimated outflow into the ambient medium). 
 The smaller problem with this model is that the optical flux rise requires the synchrotron spectral peak 
to be above optical before the 2 ks light-curve break, which is incompatible with the soft optical 
spectrum measured by UVOT at 1 ks. 
The bigger problem is the inability of the jet model to yield different decays at different frequencies, 
as measured for the optical and X-ray light-curves of afterglow 090510 after the 2 ks break 
(when power-law decay indices differ by $\alpha_{x2} - \alpha_{o2} \simeq 1$). 
 
%\vspace*{2mm}
{\sl Structured outflow}.
 In order for the emission from different parts of the outflow to be dominant at only one frequency
(above the peak energy of the synchrotron spectrum), the slope of the synchrotron spectrum must not 
be the same. Therefore, the optical and X-ray flux decays could be decoupled if the forward-shock 
accelerates electrons to a power-law distribution with energy whose index $p$ is a function of e.g. 
the shock's local Lorentz factor.
 A potential difficulty for this model is that the decoupling of the light-curve decays at two different
frequencies is not sustainable for a long time, with one part of the outflow becoming eventually dominant 
at both frequencies, due to the continuous decrease of the synchrotron peak frequency.
 In the particular case of afterglow 090510, a "top-hat" jet model does not work because the electron 
indices required for each outflow region (by the optical and X-ray flux decays which they should produce)
are not sufficiently different. 

%\vspace*{2mm}
{\sl Synchrotron self-Compton}.
 Some X-ray afterglows display a light-curve plateau at 1--10 ks whose end is chromatic and not due to a 
spectral break crossing the X-rays, which rules out a common origin of the optical and X-ray afterglow 
emissions in at least some afterglows. This may suggest that the X-ray light-curve plateau of afterglow 
090510 may not be the forward-shock emission.
 However, even if only the X-ray measurements after the plateau are attributed to the forward-shock,
a successful synchrotron forward-shock emission model cannot be found for afterglow 090510.
The problem is that the X-ray flux extrapolated back in time to 100 s and in frequency to GeV is smaller 
by a factor of a few than the GeV flux measured at 100 s. This shows that the lower and higher frequency 
emissions of GRB afterglows 090510 cannot arise from the same spectral component, which prompted us to 
consider the synchrotron self-Compton (SsC) forward-shock model, where the lower energy (optical) afterglow 
emission is synchrotron and the higher frequency (GeV) emission is up-scattering of the synchrotron photons.

 The only remaining ingredient is a reason for the achromatic light-curve break seen at 2 ks. Because
achromatic breaks must arise from the dynamics of the outflow and because a jet-break would yield a synchrotron
optical flux decay $F_{o2} \propto t^{-p}$ that is too steep, we have considered a wide outflow
for which the achromatic 2 ks break arises from cessation of energy injection in the forward-shock.

 Analytically, it is expected that the inverse-Compton flux from an adiabatic forward shock decays approximately 
as $F_x \propto t^{-(p-1.3)}$ for a homogeneous ambient medium and as $F_x \propto t^{-p}$ ($t^{-(p-1)}$) for 
a wind medium, at frequencies below (above) the cooling frequency of the inverse-Compton spectrum. 
 Therefore, if the ambient medium were homogeneous, the expected flux decay would be too slow compared to 
the post-break, fast X-ray light-curve decay of afterglow 090510. For a homogeneous medium, the SsC model 
also has difficulties in accounting for the relative intensity of the optical and X-ray fluxes.
However, the multiband fluxes of afterglow 090510, as well as their measured decays, are well accommodated 
by a wind-like medium. The best-fit obtained with this model is not without flaws (GeV flux decay is slightly 
too slow, post-break optical flux decay is slightly too fast) and some of its features are a bit extreme 
(jet energy is higher by a factor 10 than for other afterglows, wind is denser by a factor 10 than the 
average measured for Galactic Wolf-Rayet stars).

 For all the numerical calculations whose results were presented above, we find that the forward-shock 
emission is brighter than the reverse-shock's at all frequencies of interest (optical, X-ray, GeV) if 
the microphysical parameters for electron and magnetic field energy are the {\sl same} behind both shocks.
(This conclusion applies mostly to the cases where there is energy injection into the blast-wave,
i.e. when there is a reverse-shock during the afterglow phase). Modelling the multiwavelength emission
of afterglow 090510 with only the reverse-shock emission has failed to yield acceptable fits, either
because the synchrotron model emission is well below the measured GeV flux or because it exceeds the
measured optical flux when the GeV data are fit with inverse-Compton emission. However, the reverse-shock
emission depends on parameters quantifying the history of ejecta release, thus the parameter space
to be investigated is larger than for the forward-shock, thus we do not rule out a possible origin of
the GRB afterglow 090510 in the reverse-shock energizing some incoming ejecta. 

%\vspace*{2mm}
{\sl GRB 090510: a short-burst from a massive star}.
 If the SsC + energy-injection is the correct explanation for afterglow 090510, then the radial 
stratification of the ambient medium is incompatible with this short burst originating from the merger 
of two compact stars. Furthermore, for the best-fit outflow parameters (kinetic energy per solid angle 
and ambient density), the lack of a jet-break in the synchrotron optical light-curve requires a jet energy 
above $10^{52}$ erg, which may exceed what the merger of two compact stars can deposit into an 
ultrarelativistic outflow. 
%For the best-fit parameters, the medium density at the location where the afterglow emission occurs 
%($n = 10^5\, {\rm cm}^{-3}$ at the radius of $\sim 10^{16}$ cm corresponding to an observer time of 100 s, 
%and $n = 100\, {\rm cm}^{-3}$ at radius $\sim 3.10^{17}$ cm for an observer time 10 ks) is incompatible 
%with a progenitor that has escaped the host galaxy. 

 Instead, the required type of medium and outflow energetics indicate that GRB 090510 is a short burst 
arising from a massive star. If so, it must have occurred within the host galaxy, yet the center of that
host (at $z=0.9$, identified by Rau et al 2009) is at 9 kpc (in projection) from the afterglow position 
(McBreen et al 2010). That is very puzzling, yet not unprecedented: GRB 070125 is a long-burst (hence, 
it is likely to have originated also from the core-collapse of a massive star) which, probably, occurred 
in a young star-forming cluster located far ($> 27$ kpc) from the host center and well into its halo 
(Cenko et al 2008). Alternatively, GRB 090510 occurred within its host galaxy but that host is too dim 
to be detected (although its redshift cannot be higher than $z=1.5$, given the detection of the afterglow 
in near-UV bands by UVOT -- Kuin et al 2009).

 A short burst arising from a massive stellar progenitor is not the traditional view about the origin of 
bursts shorter than 2 s. The usual dichotomy -- short GRBs are from binary mergers \& long GRBs
from core-collapse of massive stars -- is based partly on the correlation of burst timescale with those 
expected for the "central-engine" of the two types of progenitors (accretion timescale below 0.1 s for 
a compact binary merger, envelope infall-time of 100 s for a massive star). It is more likely that the burst 
duration is determined by the duration of the ejection of a relativistic jet by the BH left by the core 
collapse and/or by the timescale for dissipating the jet kinetic energy (to produce the burst emission), 
both of which can be shorter than the envelope infall time. That should certainly be the case if the 
core-collapse of massive stars accounts for the long-bursts that last only a few seconds, suggesting that 
the death of massive stars could also produce bursts lasting as short as 0.5 s, as for GRB 090510.
 
 Zhang et al (2009) have proposed a replacement of the above short-long GRB dichotomy with one based on 
the properties expected for each type of progenitor: type I bursts are GRBs without associated supernovae,
occurring in hosts with a low star-formation rate or having large offsets from the host center, while 
type II bursts are GRBs occurring in hosts with high star-formation rate, tracking the host light, and 
with associated supernovae or without deep limits on it. While recognizing that type I bursts are more 
likely to originate from binary mergers and type II bursts from massive stars, Zhang et al (2009) 
also suggest that some (or most) of the higher-luminosity short bursts (like GRB 090510) are of type II
and originate from massive stars. 
 The possibility that GRB 090510 is a short burst arising from a massive stellar progenitor also finds some
support in the work of Virgili et al (2011), who argue that the luminosity-redshift distribution and the
peak flux distribution of short GRBs can be explained only if a large fraction of them follow the
star-formation history and are of type II.

\end{document}